\documentstyle[sprocl]{article}

\def\Journal#1#2#3#4{{#1} {\bf #2}, #3 (#4)}

\def\NPB{{\em Nucl. Phys.} B}

\def\PRL{\em Phys. Rev. Lett.}
\def\PRD{{\em Phys. Rev.} D}

\def\be{\begin{equation}}
\def\ee{\end{equation}}
\def\bea{\begin{eqnarray}}
\def\eea{\end{eqnarray}}
\def\cx{c_x}
\def\cy{c_y}
\begin{document}

\title{QCD ON A TRANSVERSE LATTICE}

\author{ Matthias Burkardt and Sudip Seal }

\address{Department of Physics, New Mexico State University,\\ 
Las Cruces, NM 88003-0001, USA} 


\maketitle%
\abstracts{We present results from a transverse lattice study of low
lying mesons. Special emphasis is put on the issue of Lorentz invariant
energy-momentum dispersion relations for these mesons. The light-cone
wave function for the $\pi$ obtained in this framework is very close to
its asymptotic shape.}
  
\section{Introduction}
The transverse lattice \cite{bardeen} is an attempt to combine advantages 
of the light-front (LF) and lattice formulations of QCD. 
In this approach to QCD the time and one space direction (say $x^3$)
are kept continuous, while the two `transverse' directions
${\bf x}_\perp \equiv (x^1,x^2)$ are discretized.
Keeping the time and $x^3$ directions continuous has the advantage of
preserving manifest boost invariance for boosts in the $x^3$ direction.
Furthermore, since $x^\pm = x^0 \pm x^3$ also remain continuous,
this formulation still allows a canonical LF Hamiltonian approach.
On the other hand, working on a position space lattice in the transverse
direction allows one to introduce a gauge invariant cutoff on
$\perp$ momenta --- in a similar fashion as is done in Euclidean or
Hamiltonian lattice gauge theory. 

In summary, the LF formulation has the advantage of utilizing degrees of 
freedom that are very physical since many high-energy scattering observables
(such as deep-inelastic scattering cross sections) have very simple
and intuitive interpretations as equal LF-time ($x^+$) correlation functions.
Using a gauge invariant (position space-) lattice cutoff in the $\perp$ direction within the Lf framework has the advantage of being able to avoid 
the notorious $1/k^+$ divergences from the gauge field in LF-gauge which
plague many other Hamiltonian  LF approaches to QCD \cite{mb:korea}.

The hybrid treatment (continuous versus discrete) of the long./$\perp$ 
directions implies an analogous hybrid treatment of the long. versus $\perp$ gauge field: the long. gauge field degrees of freedom
are the non-compact $A^\mu$, while the $\perp$ gauge degrees of freedom are 
compact link-fields. Each of these degees of freedom depend on two continuous 
($x^\pm$) and two discrete (${\bf n}_\perp$) space-time variables, i.e. from
a formal point of view the canonical 
$\perp$ lattice formulation is equivalent to a 
large number of coupled $1+1$ dimensional gauge theories 
(the long. gauge fields at each ${\bf n}_\perp$) coupled to nonlinear
$\sigma$ model degrees of freedom (the link fields).

\section{The color dielectric formulation}
For a variety of reasons it is advantageous to work with $\perp$ gauge degrees
of freedom that are general matrix fields rather than $U \in SU(N_C)$.
First of all, we would like to work at a cutoff scale which is small
(in momentum scale) since only then do we have a chance to find low lying
hadrons that are simple (i.e. contain only few constituents). If one wants
to work on a very coarse lattice, it is useful to imagine introducing smeared
or averaged degrees of freedom. Upon averaging over neighboring `chains' of $SU(N_C)$ fields one obtains degrees of freedom that
while they still transform in the same way as the original $SU(N_C)$ degrees of 
freedom under gauge transformations, they are general matrix degrees of freedom
which no longer obey $U^\dagger U=1$ and $det(U)=1$. The price that one has to pay for introducing these smeared degrees of freedom are more complicated
interactions. The advantage is that low lying hadrons can be described in
a Fock expansion (this has been confirmed by calculations of the static
quark-antiquark potential \cite{mb:bob} and glueball spectra \cite{brett}).

Another important advantage of this `color-dielectric' approach is that it 
is much easier to construct a Fock expansion of states out of general linear matrix fields than out of fields that are subject to non-linear $SU(N_C)$ constraints.

In the color-dielectric approach the complexity
is shifted from the states to the Hamiltonian: In principle, there exists an exact prescription for the transformation from one set of degrees of freedom
(here $U$'s) to blocked degrees of freedom $M\equiv \sum_{av} \prod_i U_i$
\be
e^{-S_{eff.}(M)} = \int \left[dU\right] e^{-S_{can.}(U)}
\delta \left(M- \sum_{av}\prod_i U_i\right).
\ee
The problem with this prescription is that $S_{eff.}$ is not only very
difficult to determine directly, but in general also contains arbitrarily complicated interactions.

A much more practical approach towards determining the effective interaction among the link fields nonperturbatively is the use of Lorentz invariance.
This strategy has been used in a systematic study of glueball masses 
in Ref. \cite{brett},
where more details can be found regarding the effective interaction.
One starts by making the most general ansatz for the effective interaction
which is invariant under those symmetries of QCD that are not broken by the
$\perp$ lattice. This still leaves an infinite number of possible terms and
for practical reasons, only terms up to fourth order  in the fields and
only local (in the $\perp$ direction) terms have been included in the 
Ref. \cite{brett}. The coefficients of the remaining terms are then fitted
to maximize Lorentz covariance for physical observables, such as the
$Q\bar{Q}$ potential (rotational invariance!) and covariance of the
glueball dispersion relation. It should be emphasized that these are first
principle calculations in the sense that the only phenomenological input
parameter is the overall mass scale (which can for example be taken to be 
the lowest glueball mass or the string tension). The only other input that is used is the requirement of Lorentz invariance.

The numerical results from Refs. \cite{brett,mb:bob} within this approach
are very encouraging:
\begin{itemize}
\item with only a few parameters, approximate Lorentz invariance could be achieved for relatively large number of glueball dispersion relations
simultaneously \cite{brett} as well as for the $Q\bar{Q}$ potential
\item the glueball spectrum that was obtained numerically on the $\perp$
lattice for $N_C\rightarrow \infty$ is consistent with Euclidean Monte
Carlo lattice gauge theory calculations performed at finite $N_C$ and extrapolated to $N_C\rightarrow \infty$ .
\end{itemize}
For further details on these very interesting results, the reader is referred
to Refs. \cite{brett,mb:bob} and references therein.

\section{Fermions on the $\perp$ lattice}
Encouraged by the very successful calculations within the pure glue sector,
we proceeded to conduct numerical studies that include fermions \cite{mb:hala}.
In this framework, states that have meson quantum numbers consist either
of a $q$ and a $\bar{q}$ on the same transverse site with no link fields
required or of a $q$ and a $\bar{q}$ at an arbitrary $\perp$ separation
with a chain of link fields in the $\perp$ direction connecting them.
Hopping of the quarks in the $\perp$ direction is accompanied by
emission or absorption of link field quanta on the link across which the
quarks hop. For each transverse site there is a longitudinal gauge field
interaction (very similar to the interaction in $QCD_{1+1}$) which couples
to $q$ and $\bar{q}$ on that site as well as to link fields adjacent to
that site. In the color dielectric approach the link fields are also
subject to the effective interaction discussed above.

Similar to other lattice field theories with fermions, species doubling
also occurs for the $\perp$ lattice action. Of course, one main difference 
to the Euclidean formulation is that the naive $\perp$ lattice action for
fermions on the $\perp$ lattice exhibits only $2^2=4$ fold species doubling,
since only two directions are discretized. Nevertheless, although
species doubling is a less extreme phenomenon here, it is a problem that needs
to be addressed.

At this point one has several options to proceed. One obvious possibility
is to add a Wilson $r$-term of the form
\be
\delta {\cal L}_r = ar\bar{\psi}{\bf \partial}_\perp^2 \psi
\ee
to the $\perp$ lattice action. Obviously, such a term violates chiral symmetry for finite lattice spacing, but this is just a consequence of the well known
Nielson-Ninomya theorem, which states that any local and hermitian action for
lattice fermions which is chirally symmetric does necessarily exhibit species doubling.

Within the LF framework there seems to be an alternative way to eliminate
`doublers': The crucial observation is that it is possible to write down
a fermion (kinetic) mass term within this framework which is chirally invariant
(but nonlocal)
\be
\delta {\cal L}_m = \delta m ^2 \bar{\psi} \frac{\gamma^+}{i\partial_-}\psi.
\ee
Since it is possible to write down a chirally invariant mass term, it is
also possible to write down a chirally invariant r-term to remove the
doublers. \footnote{Of course, it is non-local, which is why this does not
contradict the Nielson-Ninomya theorem.} 
In Ref. \cite{mb:hala} we investigated the differences between adding a
conventional r-term and such a modified chirally invariant r-term to 
the $\perp$ lattice action. The main problem is that in the canonical
LF approach, where half the fermion degrees of freedom are eliminated 
using a constraint equation, the usual chiral transformations become 
dynamical operations and therefore the meaning of the usual
chiral symmetry becomes obscure. For further details on this issue see
Ref. \cite{mb:hala}. 

Both approaches to fermions on the $\perp$ lattice give rise to
two kind of hopping terms for the fermions: one that has the Dirac structure
of a vector coupling and which flips the helicity 
(hereafter referred to as spin-flip hopping) and one which has a scalar
Dirac structure and does not flip the helicity (hereafter referred to as r-term). The difference between conventional and modified r-term is how
coefficients in the LF Hamiltonian are related to coefficients in the 
Lagrangian. In the spirit of the color-dielectric approach we regard
both coefficients in the Hamiltonian as free parameters. 

For numerical reasons, we limited the Fock space to states where $q$ and
$\bar{q}$ are on the same $\perp$ lattice site (no link field, 2 particle 
Fock component) and states where $q$ and $\bar{q}$ are separated by one link 
(3 particle Fock component), i.e the femtoworm approximation \cite{mb:hala}.

\section{Fit of parameters}
The parameters in the Hamiltonian are:
\begin{itemize}
\item longitudinal gauge coupling
\item $r$-term coupling (hopping without spin flip)
\item spin-flip coupling (hopping with spin flip)
\item kinetic mass (2 particle sector)
\item kinetic mass (3 particle sector). The observables that we studied
showed little dependence on this parameter, so we kept it fixed at a
constituent mass.
\item link field mass. Similar to kinetic mass in 3 particle sector.
Furthermore, demanding Lorentz invariance in the pure glue sector, one finds a renormalized trajectory. This makes sense since the $\perp$ lattice
scale is unphysical. We keep the link field mass fixed at a value which yields
relatively small $\perp$ lattice spacing. 
\end{itemize}

The observables that we studied, including Lorentz invariance, show 
rather little
sensitivity \footnote{As long as we kept the other parameters floating!} 
to the precise values of the 3-particle sector masses, which we
thus keep at a value corresponding to a constituent mass (about half the $\rho$ mass). This leaves us with 4 free parameters.

As input parameter, we use the physical value of the $\rho$ mass, chiral symmetry in the sense that we demand a small $\pi$ mass and the physical 
string tension. As explained in the appendix, demanding only
Lorentz invariance would drive $G^2/m_\rho^2$ to infinity and would thus
give rise to string tensions that are inconsistent with phenomenology.

Although the dependence of physical parameters on the input parameters is
in general rather complex and non-perturbative, one can understand at a qualitative level how the input parameters influence the relevant physical
scales: The physical string tension in the longitudinal direction determines $G^2$ in physical units.

Modulo mass renormalization due to the Yukawa couplings,
the quark mass in the 2-particle Fock component and the gauge coupling are
strongly constrained by fitting the physical string tension (which determines
$G^2$ and the center of mass of the $\pi$-$\rho$ system.
This leaves us with only the $r$-term and the helicity flip hopping term
couplings as parameters to vary. As explained in the appendix, the helicity
flip term is not only responsible for $\pi$-$\rho$ splitting within our approximations but also for violations of Lorentz invariance (different
$\perp$ lattice spacings in physical units for different mesons).

Using a value of $G^2 \equiv \frac{g^2N_C}{2\pi} \approx 
0.4$ $GeV^2$  in physical 
units, as determined from the string tension in Ref. \cite{brett} (which is larger than the one previously used
\cite{mb:hala,sudip}), we were able to produce the physical $\pi$-$\rho$ splitting
with only a relatively small spin-flip coupling. This allows us to chose
the $r$-term large enough so that $r$-term hopping dominates over spin-flip
hopping and therefore violations of Lorentz invariance (as measured by comparing
quadratic terms in the dispersion relation of mesons in the $\pi$-$\rho$
sector) are only on the order of $20 \%$. The average $\perp$ lattice spacing
(from the dependence of the energy on ${\bf P}_\perp$) is found to be $a_\perp \approx 0.5 
fm$. 
For the $\pi$ wave function, we find a shape that is very close to the asymptotic shape ($\phi^\pi_{as.}(x)\propto x(1-x)$). 
\begin{figure}
%\begin{Large}
\unitlength1.cm
\begin{picture}(15,7.5)(.6,-8.5)
\includegraphics{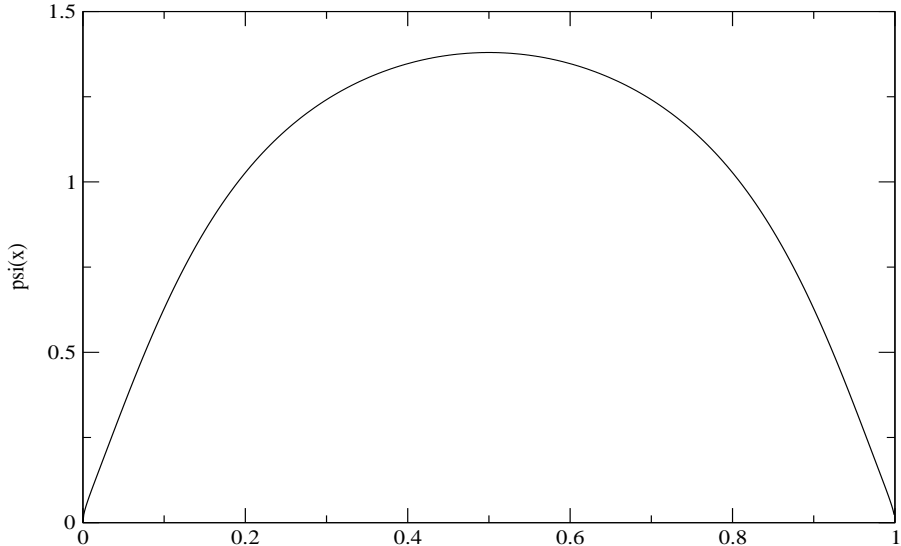}
\end{picture}
%\end{Large}
\caption{Light-cone wave function for the $\pi$ obtained on the $\perp$ lattice.
}
\label{fig:vnr}
\end{figure}
This is surprising if one consider that the lattice spacing is still relatively large and hence
the momentum scale is still very low. We should also point out that the shape
of our $\pi$ wave function disagrees with the results from Ref. \cite{sd:hd}
(the $\pi$ wave function obtained in Ref. \cite{sd:hd} is much more flat than
ours). Since the Hamiltonian and Fock space truncation
in both works are the same, the only real difference are the basis functions used to cast the Hamiltonian into
matrix form: we used continuous basis functions, while Ref. \cite{sd:hd}
uses DLCQ. However, it is not clear that this difference alone can explain the
different shape of $\pi$-wave functions.

\section*{Acknowledgments}
M.B. would like to thank H.N.-Li and W.-M. Zhang for the invitation to this interesting
workshop and S. Dalley for
many interesting and clarifying discussions about the $\perp$ lattice.
This work was supported by a grant from DOE (FG03-95ER40965) and through Jefferson Lab by contract DE-AC05-84ER40150 under which the Southeastern Universities Research Association (SURA) operates the Thomas Jefferson 
National Accelerator Facility.
 
\section*{Appendix: perturbative analysis of $\pi$-$\rho$ splitting on the
$\perp$ lattice.}
In order to gain a qualitative understanding about the interplay between different parameters in the $\perp$ lattice Hamiltonian, it is instructive
to study a simple model, where one treats the admixture of the 3-partice
Fock component to the $\pi$ and $\rho$ as a perturbation.

To $0^{th}$ order, i.e. when the coupling between 2 and 3 particle Fock component is turned off, there is no spin dependence of the interactions
and the $\pi$ and $\rho$ are degenerate. Likewise, there is no `hopping'
(i.e. $\perp$ propagation) of mesons and thus energies are independent of
${\bf k}_\perp$ giving rise to an infinite $\perp$ lattice spacing
(in physical units).

In the next order we treat the coupling between 2 and 3 particle Fock components as a perturbation (note that interactions which are diagonal in the particle
number, such as the confining interaction in the longitudinal direction are still treated non-perturbatively). There are two interactions that mix Fock sectors: hopping due to the $r$-term (without helicity flip) and hopping due to the vector coupling (with helicity flip).
In order to understand the effect of these two types of hopping, it is very useful to point out that there is no `mixing' between these two hopping terms
in the femtoworm approximation, i.e. there is a complete cancellation among the various hopping terms where the 2 to 3 transition is caused by say the $r$-term and the subsequent 3 to 2 transition is caused by the $v$-term (and the other way round). For ${\bf k}_\perp=0$ this follows trivially from the fact that the
$\perp$ lattice Hamiltonian is invariant under rotations around the $z-axis$ by multiples of $\pi/2$, giving rise to conservation of total angular momentum
modulo $4$. Since $J_z=S_z$ in the 2 particle sector, and $S_z$ can only assume the values $-1$, $0$, and $1$, $S_z$  in the 2 particle sector is conserved.
Since `mixed' hopping would change $S_z$ by one unit, this means that the sum of all contributions from mixed hopping must add up to zero.

For ${\bf k}_\perp \neq {\bf 0}$ the argument is a little more complicated, since rotations also change the direction of ${\bf k}_\perp$. However, both the $\perp$ lattice Hamiltonian as well as ${\bf k}_\perp$ are invariant under the
a sequence consisting of a rotation by $\pi$ around the $z$-axis followed by
a $\perp$ reflection on the $x$-axis and then a $\perp$ reflection on the $y$ axis $P_yP_xR_{180}$. 
As a result, $J_z$ is conserved modulo $2$, which still rules out mixing
between the $r$-term and the $v$-term.

Starting from a basis of `t Hooft eigenstates which are plane waves in the $\perp$ direction and where the $q\bar{q}$ in the 2 particle Fock component carry spins $ |\uparrow \uparrow\rangle$, 
$ |\uparrow \downarrow\rangle$, $ |\downarrow \uparrow\rangle$, 
and $ |\downarrow \downarrow\rangle$ respectively, one thus finds for the 
energy in second order perturbation theory 
\bea
H = M_0^2 &-& M_{1,r}^2\left(\begin{array}{cccc}
\cx + \cy & 0 & 0 & 0\\
0 & \cx + \cy & 0 & 0\\
0 & 0 & \cx + \cy & 0\\
0 & 0 & 0 & \cx + \cy\end{array}\right) \nonumber\\
&+& M_{1,v}^2\left(\begin{array}{cccc}
0 & 0 & 0 & \cy - \cx\\
0 & 0 & \cx + \cy & 0\\
0 &  \cx + \cy & 0 & 0\\
\cy-\cx  & 0 & 0 & 0\end{array}\right) . \label{eq:M2}
\eea
Here $M_{1,r}^2$ and $M_{1,v}^2$ are some second order perturbation theory
expressions involving matrix elements between 2 and 3 particle states that are
eigenstates of the diagonal parts of the Hamiltonian (kinetic + Coulomb),
and $c_i \equiv \cos k_i$.

Several general and important features can be read off from this result. 
First of all, and most importantly, Eq. (\ref{eq:M2}) 
Shows that the r-term gives rise to a dispersion relation with the same
$\perp$ speed of light for the $\pi$ and the $\rho$'s, 
while the vector interaction breaks that symmetry.
This observation already indicates that it may be desirable to keep the $r$-term
much larger than the spin-flip term. We will elaborate on this point below.

At $k_x=k_y=0$, the eigenstates of the above Hamiltonian are the 
$\rho_{\pm 1}$ i.e.
$|\uparrow \uparrow\rangle$ and 
$|\downarrow \downarrow\rangle$, 
with $M^2= M^2_{\pm 1}\equiv M_0^2 - M_{1,r}^2$,
the $|\rho_0\rangle\equiv 
|\uparrow \downarrow+ \downarrow \uparrow\rangle$, 
with  $M^2= M^2_{\pm 1} + M_{1,v}^2$ and
the   $|\pi\rangle\equiv 
|\uparrow \downarrow- \downarrow \uparrow\rangle$, 
with  $M^2= M^2_{\pm 1} - M_{1,v}^2$ 
For nonzero $\perp$ momenta, there will in general be mixing 
among the $\rho_{+1}$ and the $\rho_{-1}$, but not among the other states
since helicity in the 2-particle Fock sector is still conserved modulo 2.
Expanding around ${\bf k}_\perp=0$, and denoting
$\bar{M}^2\equiv M_0^2-M_{1,r}^2$
one finds to 
${\cal O}({\bf k}_\perp^2)$ the following eigenstates and eigenvalues
\be
\begin{array}{c|c|c}
\mbox{state} & M^2(0) & M^2({\bf k}_\perp^2)-M^2(0)\\ \hline 
 & & \\
\uparrow \downarrow -\downarrow \uparrow & \bar{M}^2 -M_{1,v}^2
&
M_{1,r}^2\frac{k_x^2+k_y^2}{2}+M_{1,v}^2\frac{k_x^2+k_y^2}{2}\\
& & \\
\uparrow \downarrow +\downarrow \uparrow & \bar{M}^2+M_{1,v}^2
 &
M_{1,r}^2\frac{k_x^2+k_y^2}{2}-M_{1,v}^2\frac{k_x^2+k_y^2}{2}\\
& & \\
\uparrow \uparrow -\downarrow \downarrow & \bar{M}^2
 &
{M_{1,r}^2}\frac{k_x^2+k_y^2}{2}
-{M_{1,v}^2}\frac{k_x^2-k_y^2}{2}\\
& & \\
\uparrow \uparrow +\downarrow \downarrow & \bar{M}^2
 &{M_{1,r}^2}\frac{k_x^2+k_y^2}{2}
+{M_{1,v}^2}\frac{k_x^2-k_y^2}{2}
\end{array} \label{eq:disp}.
\ee 
Eq. (\ref{eq:disp}) illustrates a fundamental dilemma that hampers any
attempt to fully restore Lorentz invariance within the femtoworm
approximation: $M_{1,v}^2$ not only governs the splitting between 
the $\pi$ and the (h=0) $\rho$ but is also responsible for violations
of Lorentz invariance among the different helicity states: If one
determines the $\perp$ lattice spacing in physical units for each 
meson separately, by demanding that the
$\perp$ speed of light equals $1$, one finds for example
\bea
\left. \frac{1}{a^2_\perp} \right|_\pi &=& M_{1,r}^2 + M_{1,v}^2
\nonumber\\
\left. \frac{1}{a^2_\perp} \right|_{\rho_0} &=& M_{1,r}^2 - M_{1,v}^2
\eea
i.e. increasing the $\pi-\rho$ splitting is typically accompanied by
an increase in Lorentz invariance violation
\be
\left. \frac{1}{a^2_\perp} \right|_\pi -
\left. \frac{1}{a^2_\perp} \right|_\rho = M_{\rho_0}^2 - M_\pi^2 .
\ee
For the $\rho_{\pm 1}$ the breaking is of a similar scale, plus one
also observes an anisotropy in the dispersion relation on the same 
scale.

Therefore, in order to avoid a large breaking of Lorentz invariance,
it will be necessary that 
\be
M_{r,1}^2 \gg M_{\rho_0}^2 - M_\pi^2.
\ee
If one keeps the $\pi-\rho$ splitting fixed at its physical value then
there are two ways to achieve this condition. One possibility is to simply
increase the Yukawa coupling that appears in the $r$-term. 
This increase of the $r$-term tends to decrease the $\perp$ lattice spacing
for both $\pi$ and $\rho$'s and in order to achieve satisfactory
Lorentz invariance (in the sense of uniform $\perp$ lattice spacings)
one needs to make the lattice spacing smaller than the Compton wavelength of 
the $\rho$ meson.
However, one cannot make the $r$-term coupling arbitrarily 
large because at some point there occurs an instability (tachyonic $M^2$!).
Such instabilities for large coupling are common in the LF formulation
of models with Yukawa coupling and might be 
related to a phase transition (similar to the phase transition in $\phi^4$ theory that occurs as the coupling is increased).

Fortunately, there exists another possibility to make these matrix 
elements large, without increasing the Yukawa couplings. This derives 
from the fact that the hopping interactions are proportional to $\left(\frac{1}{x} \pm \frac{1}{x^\prime}\right)\frac{1}{x-x^\prime}$, where $x$ ($x^\prime$) are the momenta of the active quark before(after) the hopping. Because of the singularity as $x$, $x^\prime \rightarrow 0$, matrix elements of the hopping terms are greatly enhanced if
the unperturbed wave functions are large near $x=0$ and $x=1$. 
Since the unperturbed wave functions in the 2 particle Fock component vanish
like $x^\beta$ near $x=0$, where $\beta \propto \frac{G}{m_q}$,
matrix elements of the hopping interaction become very large when
one makes $\frac{G^2}{m_q^2}$ very large.

Therefore, the larger one chooses $\frac{G^2}{m^2}$, the more one restores Lorentz invariance of the $\pi$ and $\rho$ dispersion relations because one
can keep the $\pi$-$\rho$ splitting fixed while decreasing the coupling of the
spin flip interaction. At the same time, keeping the $r$-term interaction
fixed one increases the dominance of the $r$-term contribution in $\frac{1}{a^2}$ and thus not only reduces the lattice spacing in physical units, but also obtains dispersion relations for the $\pi$ and the $\rho$'s that look
more and more similar --- as demanded by Lorentz invariance.
Unfortunately, we are not completely free to pick whatever value of
$\frac{G^2}{m^2_q}$ we like because $m_q^2$ and $G^2$ are largely fixed by the
center of mass in the $\pi$-$\rho$ system as well as by fitting the physical
string tension in the pure glue sector.


\section*{References}
\begin{thebibliography}{99}
\bibitem{bardeen} W.A. Bardeen, R.B. Pearson, and E. Rabinovici,
\Journal{\PRD}{21}{1037}{1980}.
\bibitem{mb:korea} M. Burkardt, invited talk given at ``11th International 
Light-Cone School and Workshop'', Eds. C. Ji and D.-P. Min,
hep-th/9908195. 
\bibitem{mb:bob} M. Burkardt and B. Klindworth, 
\Journal{\PRD}{55}{1001}{1997};
hep-ph/9809283.
\bibitem{brett} S. Dalley and B. vande Sande, 
\Journal{\PRD}{59}{065008}{1999}; \Journal{\PRL}{82}{1088}{1999};
\Journal{\NPB (Proc. Suppl.)}{83}{116}{2000};
\Journal{\PRD}{62}{014507}{2000}.
\bibitem{mb:hala} M. Burkardt and H. El-Khozondar, 
\Journal{\PRD}{60}{054504}{1999}.
\bibitem{sd:hd} S. Dalley, \Journal{\NPB (Proc. Suppl.)}{90}{227}{2000}.
\bibitem{sudip} S.K. Seal and M. Burkardt, 
\Journal{\NPB (Proc. Suppl.)}{90}{233}{2000}.
\end{thebibliography}
\end{document}